\documentclass[preprint,numbers,sort&compress,1p,10pt]{elsarticle}
\usepackage[T1]{fontenc}
\usepackage[latin1]{inputenc}
\usepackage{graphicx}
\usepackage{amssymb}
\usepackage{rotating}
\usepackage{epsfig}

\newcommand{\ie}{{\em i.e.}}

\newcommand{\etal}{{\em et al.}}

\journal{Comp. Phys. Comm.}

\begin{document}
\begin{frontmatter}
\title{On the existence of helical structures during the collapse of flexible homopolymers: A Wang-Landau study}

 \author[ust]{Sid Ahmed ~Sabeur\corref{cor1}}
 \ead{aminesabeur@yahoo.fr}

 \author[ust]{Mounira ~Bouarkat}

 \author[ust]{Fatima ~Hamdache}

 \cortext[cor1]{Corresponding author.}
 \address[ust]{Laboratoire de Physique des Plasmas, des Matériaux Conducteurs et de leurs Applications,
 Département de Physique, Faculté des Sciences, USTOMB, BP 1505 El M'naouer, Oran, Algeria}

\begin{abstract}
In this work, we report our results on the phase transition of a flexible homopolymer from a stretched chain
to a compact globule. The Wang-landau method is used to study the thermodynamic properties of a
the chain up to 512 monomers. We believe that the peak in the specific heat at
low temperature $T\approx 0.05$ for small chain sizes $N<100$ is a clear
evidence of the existence of metastable helical structures observed in a previous study.
\end{abstract}


\end{frontmatter}

\section{Introduction}\label{sec1}
In recent years, designing functional polymers at nanometers scales has gained increasing interest\cite{Glotzer}.
However, it is challenging to control structures that have the ability to undergo cooperative transitions between
random conformations (globules) and ordered conformations ($\alpha-$helices and $\beta-$sheets). There has been a spate of theoretical studies
dealing with this issue, commonly achieved using Monte Carlo and molecular dynamics simulations\cite{dawson1,dawson2,dawson3,Frisch2}. These works have offered rewarding insights for understanding the dynamics and the phase transition of polymers and there is still a big interest in investigating them until today\cite{Landau5}.
In a previous work, we have found that flexible homopolymers spontaneously develop helical order during the process
of collapsing from an initially stretched conformation. We have also demonstrated that the helices are long lived transient states at low temperatures\cite{Sabeur1}, but with "borrowing" energy from their surrounding solvent particles, they overcome the energetic barrier and collapse into a stable globule (Fig. 1 shows typical configurations in the process of a homopolymer collapse).
Here, we extend our study to the thermodynamic behavior of a flexible homopolymer using the Wang-Landau method\cite{Landau1}. Our main goal is to confirm the existence of the helical structures during the collapse, and to characterize the temperature regions corresponding to those structures.
The paper is organized as follows. In Sec. II we introduce the model. In Sec. III the simulation method
and the parameters used to quantify the thermodynamic properties are presented. The simulations results are
summarized in Sec. IV. Finally, conclusions and appropriate acknowledgments are given in Sec. V.

\begin{figure}[t]
\centerline{\includegraphics[scale=0.3]{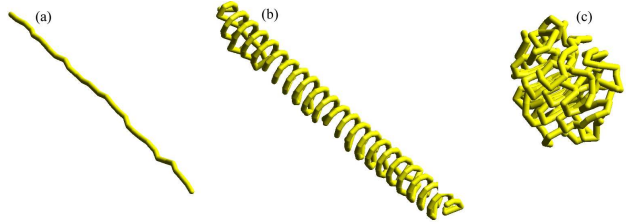}}
\caption{\label{fig1}
Typical configurations observed during the collapse of a flexible homopolymer
from an initially stretched chain at low temperatures. a) Initial stretched
structure, b) metastable helical structure,
c) final stable globular structure.}
\end{figure}

\section{Model}
In the simulation, we have used a simple bead-spring model. The polymer chain consists of $N$
identical monomers where each monomer position varies continuously in three dimensions.
There are two types of interactions: a harmonic interaction between adjacent monomers

\begin{equation}
\label{eq:vbond}
 U_{\mbox{\tiny bond}}(r)=a(r-r_0)^2,
\end{equation}

with $a=100\epsilon/\sigma^2$ and $r_0=0.85\sigma$

and a truncated Lennard-Jones potential between non adjacent monomers

\begin{equation}
\label{eq:vlj}
V_{\mbox{\tiny LJ}}(r) = \left\{
\begin{array}{l}
4 \epsilon [ (\sigma/r)^{12} - (\sigma/r)^6 + c_0 ]
\; \mbox{for $r<2.5 \sigma$}\\
0 \quad \mbox{otherwise},
\end{array}
\right.
\end{equation}

where the constant $c_0$ is chosen such that the potential is continuous everywhere.
Dimensionless units are used during the simulation and are defined in terms
of the bead size $\sigma$ and the Lennard-Jones energy $\epsilon$.

\section{Method}
We have implemented the Wang-Landau algorithm to study the thermodynamic behavior
of the flexible homo-polymer\cite{Landau6}. This method is a temperature independent
Monte Carlo technique for exploring the energy landscape. By calculating the density of states $g(E)$, where $E$
is the energy of the polymer chain, any thermodynamic observable $A$
can be obtained with one single simulation for a wide range of temperatures through
the canonical average

\begin{equation}
\label{eq:avr}
\langle A \rangle_T =\frac{\sum_E\langle A \rangle_E g(E)e^{-E/k_BT}
}{Z}
\end{equation}

Where $Z$ is the partition function and can be determined by

\begin{equation}
Z =\sum_E g(E)e^{-E/k_BT}
\end{equation}

Using Eq.~\ref{eq:avr}, one can calculate the internal energy $\langle E \rangle_T $ of the homopolymer

\begin{equation}
\langle E \rangle_T =\frac{E g(E)e^{-E/k_BT}
}{Z} \label{eq:A}
\end{equation}

and thus the specific heat $C_v$

\begin{equation}
\label{eq:cv}
  C_v =\frac{\langle E^2 \rangle_T -\langle E \rangle_T^2}{k_BT^2}
\end{equation}

Initially, we start the simulation by taking a stretched configuration for the homopolymer chain.
We set the density of states $g(E)=1.0$. Then, we proceed by generating
trial states, using appropriate "moves", and accepting them with transition
probability

\begin{equation}
p(E_i\rightarrow E_f )= min \left(\frac{g(E_i)}{g(E_f)},1 \right)
\end{equation}

where $E_i$ is the initial energy and $E_f$ is the energy of the trial
state.

Each Monte Carlo sweep in the simulation consists of the following moves: $N$ diffusion
moves, one reptation move, one single-bead crankshaft move and one pivot move. The moves
are chosen identical to those of reference\cite{Landau4}.

If the trial move is accepted, the density of states is multiplied
by a modification factor f , \ie  $g(E_f)\rightarrow g(E_f)\times f$ , and
a histogram $H(E)$ is also updated, $H(E_f)\rightarrow H(E_f)+1$. If the
trial state is rejected, the same procedure is followed for the
initial energy $E_i$. This process is repeated until the histogram
H(E) is sufficiently flat, at which point the histogram is reset
to zero, the modification factor is reduced by $f \rightarrow \sqrt f$ , and
the random walk continues. The initial value of the modification factor is set to $f = e^1$
and is eventually reduced to a value which is $f_{\mbox{final}} \approx 1$.
The flatness of $H(E)$ is defined as its minimum divided by its average, and in this study, the
flatness criterion is set to 0.6. To prevent overflow during the simulations, the logarithm of
the density of states $ln[g(E)]$ is used instead of $g(E)$. More details about the Wang-Landau algorithm
can be found in reference\cite{Landau1}

In this study we have defined the energy range per monomer $E/N=[-3.75,3.0]$. Due to overflows in the computation,
the minimum energy boundary is far from the real $E_{\tiny \mbox{min}}/N=-7.75\epsilon$, found in a previous work\cite{Sabeur1}.
Pushing the minimum energy to the lower limit needs more computational efforts and has to
be explored in the future. The bin resolution is taken $dE=0.1$.
All the results obtained represent averages over 20 independent runs.

\section{Simulation results}
\begin{figure}[t]
\centerline{\includegraphics[scale=0.7]{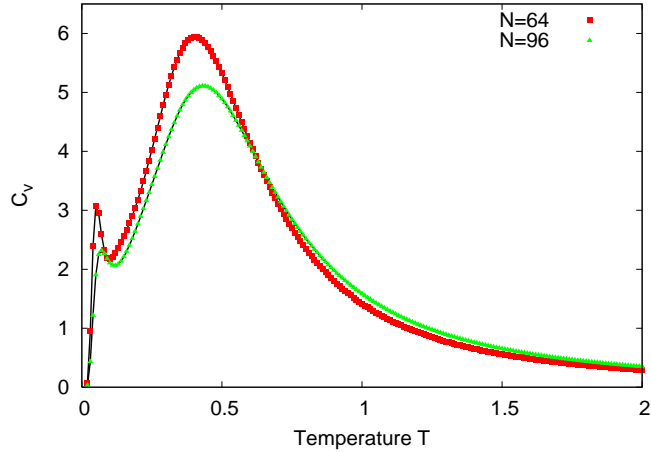}}
\caption{Specific heat versus temperature for chain lengths $N=64,96$. Two
  features appear: a large peak at temperature $T\approx 0.4$ representing
the coil-globule transition and a sharp peak at low temperatures around
$T\approx0.05$ corresponding to the formation of helical structures.}
\label{fig2}
\end{figure}

The temperature dependence of the specific heat for chain lengths smaller then
$N=100$  are shown in Fig.~\ref{fig2}. The large peak around $T\approx0.4$
corresponds to the coil-globule transition.

At lower temperatures, for $N=64$, a sharp peak forms
around $T\approx0.05$, indicating another transition.  We believe that this
transition corresponds to the formation of helical structures. The peak becomes smaller
for $N=96$.

\begin{figure}[h]
\centerline{\includegraphics[scale=0.7]{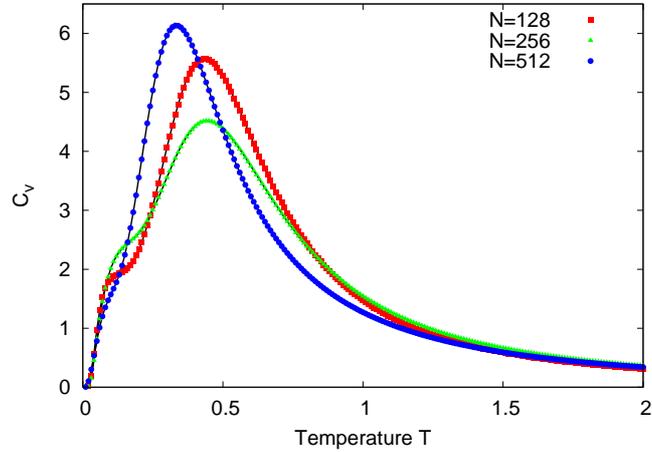}}
\caption{Specific heat versus temperature for chain lengths $N=128,256$ and
  $512$. These chain lengths offer two features: a large peak in $C_v$
  indicative of the coil-globule transition and a shoulder indicative of
  rapidly vanishing metastable states.}

\label{fig3}
\end{figure}

Fig.~\ref{fig3} also shows the temperature dependence of the specific heat but
for chain sizes greater than $N=100$. At low temperature, for chain lengths
$N=128$ and $N=256$, we observed large shoulders. This can be explained by the
fact that it is difficult for helical structures to form correctly for longer chain sizes.
Those shoulders are rapidly vanishing for greater chain sizes ($N=512$)
because of the metastable nature of the helical structures.

\section{Conclusions}
We have studied the thermodynamic behavior of a flexible homopolymer using
the Wang-Landau method for chain sizes up to $N=512$. Our results show two
kind of transitions. The coil-globule transition is observed around
the temperature $T\approx 0.4$, for all the chain
sizes studied . At low temperature $T\approx0.05$, a second transition appears and
depends on the homopolymer chain size. For small chain sizes
$N<100$, a sharp peak appear corresponding to a long lived helical structures.
However, for $N>100$ we observe a shoulder that become small as the chain size
increases. We believe that for infinite chain sizes, this transition
vanishes and that the formation of helical structures occurs only for finite
chain sizes. Although, the potential model used for the flexible homopolymer
is different from the one used by Seaton \etal \cite {Landau3}, our results
tend to have some similarities.

\section*{Acknowledgments}
This work has been supported by the Algerian Ministry of Higher Education and
Scientific Research, within the frame work of projects CNEPRU-D01920001054.

\bibliographystyle{cpc}
\bibliography{biblio}

\begin{thebibliography}{10}

\bibitem{Glotzer}
{Trung Dac Nguyen and Sharon C. Glotzer},
\newblock small {\bf 5(18)} (2009) 2092.

\bibitem{dawson1}
{E.G Timoshenko, Yu. A. Kuznetsov and K.A. Dawson},
\newblock J. Chem. Phys. {\bf 102} (1995) 1816.

\bibitem{dawson2}
{E.G Timoshenko, Yu. A. Kuznetsov and K.A. Dawson},
\newblock J. Chem. Phys. {\bf 104} (1996) 3338.

\bibitem{dawson3}
{E.G Timoshenko, Yu. A. Kuznetsov and K.A. Dawson},
\newblock Phys. Rev. E. {\bf 51} (1996) 3381.

\bibitem{Frisch2}
{T. Frisch and A. Verga},
\newblock Phys. Rev. E {\bf 66} (2002) 041807.

\bibitem{Landau5}
{D. T. Seaton, T. W\"ust and D. P. Landau},
\newblock Phys. Rev. E {\bf 81} (2010) 011802.

\bibitem{Sabeur1}
{S. A. Sabeur, F. Hamdache and F. Schmid},
\newblock Phys. Rev. E {\bf 77} (2008) 020802.

\bibitem{Landau1}
{F. Wang and D. P. Landau},
\newblock Phys. Rev. E {\bf 64} (2001) 056101.

\bibitem{Landau6}
{F. Wang and D. P. Landau},
\newblock Phys. Rev. Lett {\bf 86(10)} (2001) 2050.

\bibitem{Landau4}
{D. T. Seaton, S. J. Mitchell, D. P. Landau},
\newblock Braz. J. Phys. {\bf 38} (2007) 1.

\bibitem{Landau3}
{D.T. Seaton, T. Wüst and D.P. Landau},
\newblock Comput. Phys. Comm. {\bf 180} (2009) 587.

\end{thebibliography}

\end{document}